\documentclass[conference]{IEEEtran}

\usepackage{graphicx} 
\usepackage[nospace,sort,adjust,compress]{cite} 
\usepackage{hyperref}
\usepackage{url} 
\usepackage{breakurl} 
\usepackage{booktabs} 
\usepackage[small]{caption} 
\usepackage[font=small]{subcaption} 

\begin{document}

\title{Security for Smart Mobile Networks:\\The NEMESYS Approach}

\author{
	\IEEEauthorblockN{Erol Gelenbe\IEEEauthorrefmark{1}, Gokce Gorbil\IEEEauthorrefmark{1}, Dimitrios Tzovaras\IEEEauthorrefmark{2}, Steffen Liebergeld\IEEEauthorrefmark{3},\\
	David Garcia\IEEEauthorrefmark{4}, Madalina Baltatu\IEEEauthorrefmark{5} and George Lyberopoulos\IEEEauthorrefmark{6}}
	\IEEEauthorblockA{\IEEEauthorrefmark{1}Imperial College London, Dept. of Electrical and Electronic Engineering,\\
	SW7 2AZ London, United Kingdom\\
	Email: \{e.gelenbe, g.gorbil\}@imperial.ac.uk}
	\IEEEauthorblockA{\IEEEauthorrefmark{2}Centre for Research and Technology Hellas, Information Technologies Institute,\\
	57001 Thessaloniki, Greece}
	\IEEEauthorblockA{\IEEEauthorrefmark{3}Technical University of Berlin, 10587 Berlin, Germany}
	\IEEEauthorblockA{\IEEEauthorrefmark{4}Hispasec Sistemas S.L., 29590 Campanillas (Malaga), Spain}
	\IEEEauthorblockA{\IEEEauthorrefmark{5}Telecom Italia IT, 20123 Milan, Italy}
	\IEEEauthorblockA{\IEEEauthorrefmark{6}COSMOTE - Mobile Telecommunications S.A., 15124 Maroussi, Greece}
}

\maketitle


\begin{abstract}
The growing popularity of smart mobile devices such as smartphones and tablets has made them an attractive target for cyber-criminals, resulting in a rapidly growing and evolving mobile threat as attackers experiment with new business models by targeting mobile users. With the emergence of the first large-scale mobile botnets, the core network has also become vulnerable to distributed denial-of-service attacks such as the signaling attack. Furthermore, complementary access methods such as Wi-Fi and femtocells introduce additional vulnerabilities for the mobile users as well as the core network. In this paper, we present the NEMESYS approach to smart mobile network security. The goal of the NEMESYS project is to develop novel security technologies for seamless service provisioning in the smart mobile ecosystem, and to improve mobile network security through a better understanding of the threat landscape. To this purpose, NEMESYS will collect and analyze information about the nature of cyber-attacks targeting smart mobile devices and the core network so that appropriate counter-measures can be taken. We are developing a data collection infrastructure that incorporates virtualized mobile honeypots and honeyclients in order to gather, detect and provide early warning of mobile attacks and understand the modus operandi of cyber-criminals that target mobile devices. By correlating the extracted information with known attack patterns from wireline networks, we plan to reveal and identify the possible shift in the way that cyber-criminals launch attacks against smart mobile devices.
\end{abstract}



\section{Introduction}
\label{sec:intro}

Smartphones, tablets, and similar smart mobile devices have been greatly successful as personal communication and computing devices owing to their always-on connectivity, mobility, usability, and operating systems that enable a vast market of applications tailored for the needs of mobile users. Smart devices are undeniably taking over the mobile device market, and this trend is not expected to slow down in the near future. Table \ref{tab:deviceShipments} gives the number of mobile devices shipped worldwide in 2012 based on data from Canalys~\cite{bib:canalysMobileMarket13}. The data shows that smartphones and tablets represented $42\%$ of all mobile devices shipped in 2012, and their combined market share is expected to grow to $66\%$ by 2016. The significance of these numbers is highlighted by the fact that among all consumer devices, only the television has had such a fast market penetration rate in the USA~\cite{bib:degustaTech12}. Furthermore, while smartphones represented only $18\%$ of total global mobile phones in use in 2012, they were responsible for $92\%$ of total global mobile phone traffic~\cite{bib:ciscoMobileForecast13}. With both the device market share and the amount of data traffic due to smart devices expected to continue to grow significantly, their importance is evident in the current and future mobile landscape.

\setlength{\tabcolsep}{8pt}

\begin{table}[tbp]
	\caption{The number of mobile devices shipped worldwide in 2012 and forecast for 2016, in millions of units, according to Canalys (Feb. 2013) \cite{bib:canalysMobileMarket13}.}
	\label{tab:deviceShipments}
	\centering
	\begin{tabular}{r l l l}
		\toprule
		Device type		& 2012 shipments & 2016 shipments & 2012--2016 CAGR\\
		\midrule
		Basic phone		& $122.0$ & $58.0$ & $-17.0\%$\\
		Feature phone	& $770.8$ & $660.9$ & $-3.8\%$\\
		Smartphone		& $694.8$ & $1,342.5$ & $17.9\%$\\
		Tablet			& $114.6$ & $383.5$ & $35.3\%$\\
		Notebook			& $215.7$ & $169.1$ & $-5.9\%$\\
		Netbook			& $18.3$ & $0.3$ & $-64.2\%$\\
		\midrule
		Total 			& $1,936.2$ & $2,614.2$ & $7.8\%$\\
		\bottomrule
	\end{tabular}
\end{table}

\begin{figure*}[tbp]
	\centering
	\begin{subfigure}[b]{0.45\linewidth}
		\centering
		\includegraphics[width=\linewidth, trim = 0cm -1.6cm 0cm 0cm]{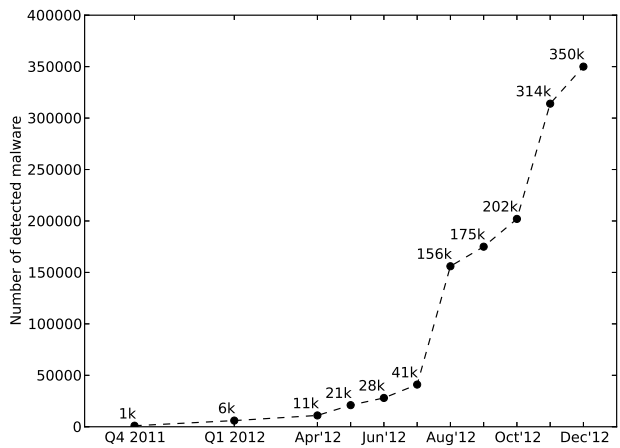}
		\caption{The number of detected Android malware in 2012. The data shows a rapid increase in malware, especially in Q3-Q4 2012, which was due to aggressive adware.}
		\label{fig:androidMalware}
	\end{subfigure} \hspace{0.5cm}
	\begin{subfigure}[b]{0.45\linewidth}
		\centering
		\includegraphics[width=\linewidth]{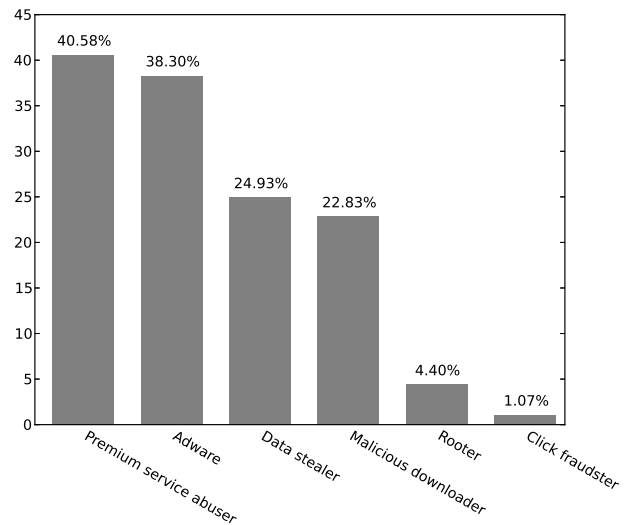}
		\caption{The distribution of malware types for the top ten Android malware families in 2012}
		\label{fig:malwareTypes}
	\end{subfigure}
	\caption{Android malware in 2012, based on data provided by TrendLabs~\cite{bib:trendmicroMobileThreats12}}
	\label{fig:malware}
\end{figure*}

Smart devices increasingly store and provide access to a range of personal, corporate, financial and security-related data. While the availability and ease of access to such data via different types of network connections (e.g.~\mbox{Wi-Fi}, cellular), the ability to download, install and use mobile applications, and the practicality of using networked services while on the go make the smart mobile device an inseparable companion to the modern man, they also provide the perfect breeding ground for malicious software. Cyber-criminals have not had continuous access to such varied sources of personal and financial data before the advent of the smartphone. Increasing use of mobile payments presents a new area to exploit, and attackers have already devised covert ways to make direct financial gain from the mobile users, e.g. via premium calls and SMS~\cite{bib:feltSurveyMobileMalware11}. Furthermore, the use of multiple communication technologies in smart devices allows attackers to cross service boundaries. For example, an attack carried out over \mbox{Wi-Fi} would allow the attacker to launch attacks over the mobile phone network. In addition to exposing the mobile user to heterogeneous attack vectors, the increasing use of complementary access to the mobile core network, via \mbox{Wi-Fi} and femtocells, introduces new vulnerabilities to the core network~\cite{bib:goldeFemtocell12}. The amount of data carried over complementary access should not be underestimated; according to Cisco~\cite{bib:ciscoMobileForecast13}, $429$ petabytes per month of global mobile data was offloaded onto the fixed network through \mbox{Wi-Fi} or femtocell radio access, which represents $33\%$ of total mobile traffic in 2012. Cisco also estimates that mobile data offload will reach $9.6$ exabytes per month by 2017, which will be $46\%$ of total mobile traffic.

Smart mobile devices are also increasingly at the center of security systems for managing small or large emergencies in built environments, or during sports or entertainment events~\cite{bib:gelenbeICCCN12,bib:gelenbeWuSimEvac12}, and they are also used increasingly for online search of difficult-to-get sensitive information~\cite{bib:gelenbeSearch10,bib:abdelrahmanSearch13}. Thus such technologies will necessarily be targeted and may be breached in conjunction with other physical or cyber attacks, as a means of disrupting safety and confidentiality of individuals and emergency responders~\cite{bib:gorbilANT11,bib:gorbilISCIS11,bib:gorbilPernem13}.

In order to address the growing mobile threat, there is an urgent need to detect, analyze and understand the new vulnerabilities and threats in the smart mobile ecosystem, which are a result of the evolution of mobile networks and smart devices, the changing way users interact with technology, the popularity of smart devices, and the heterogeneity of the wireless interfaces, supported platforms and offered services. In order to advance in the fast moving field of cyber-security and to counter existing and potential mobile threats, we need to be proactive and work on predicting threats and vulnerabilities to build our defenses before threats materialize. To this purpose, the EU FP7 NEMESYS project\footnote{\url{http://www.nemesys-project.eu/nemesys/}} will develop a novel security framework to gather and analyze information about the nature of cyber-attacks targeting mobile devices and the mobile core network, as well as identify and predict abnormal behaviour observed on smart mobile devices and the mobile network so that appropriate countermeasures can be taken. We aim to understand the modus operandi of cyber-criminals, and to identify and reveal the possible shift in the way they launch attacks against mobile devices through root cause analysis and correlation of new findings with known patterns of attacks on wireline networks.

\section{The Current Mobile Threat Landscape}
\label{sec:landscape}

Smart devices are open to both traditional and mobile-specific threats due to the multiple roles smart mobile devices play and the heterogeneity of mobile communication technologies and networked services~\cite{bib:becherMobileSecurity11}. Among the traditional threats that smart mobile devices face, we include physical attacks that require physical access to the device, device-independent attacks such as eavesdropping on the wireless medium or man-in-the-middle attacks, e-mail-based spam and phishing, and IP-based attacks. Current IP-based attacks encountered on mobile devices~\cite{bib:wahlischMobileHoneypot13} have been found to be largely similar to those on non-mobile devices~\cite{bib:gelenbeLoukasDoS07,bib:gelenbeSelfAware09}, but we are more interested in the traits of attacks that are tailored specifically for mobile devices. With the growing popularity of smart devices, mobile-specific threats have evolved from SMS/MMS-based denial-of-service (DoS) attacks~\cite{bib:traynorSMSAttack09,bib:mullinerSMSAttack11} to more sophisticated attacks that usually come in the form of malware and target both the core network and the mobile users. The ability of smart devices to install and run applications not only from official markets but also from unknown sources exposes them to malware~\cite{bib:feltSurveyMobileMalware11,bib:zhouAndroidMalware12}, and while the mobile malware threat is not new~\cite{bib:dagonMobileVirus04}, it is clearly evolving and growing as attackers experiment with new business models by targeting mobile users~\cite{bib:lookoutMobileSecurity12,bib:kasperskyMalwareEvolution12}. We next provide a taxonomy of mobile malware based on behavioral classification, and present an overview of malware detection techniques.

\subsection{Mobile Malware}

Android has been the most targeted mobile platform in 2012, with almost $99\%$ of all encountered mobile malware being designed for Android~\cite{bib:kasperskyStatistics12}. The number of malicious Android applications detected by Kaspersky Lab in 2012 was more than $35,000$, which reflects a six-fold increase from 2011~\cite{bib:kasperskyMalwareEvolution12}. Figure \ref{fig:androidMalware} shows the rapid growth of Android malware in 2012 based on data from TrendLabs~\cite{bib:trendmicroMobileThreats12}, which shows the significance of the growing mobile malware threat. 2012 has also seen the emergence of the first mobile botnets~\cite{bib:kasperskyStatistics12}. A botnet is a collection of Internet-connected devices acting together to perform tasks, often under the control of a command and control server. In wireline networks, malicious botnets are used to generate various forms of spam, phishing, and distributed denial-of-service (DDoS) attacks. Mobile botnets extend such capability to cellular networks, give cyber-criminals the advantages of control and adaptability, and pose a significant threat to the mobile core network as they could be used to launch debilitating signaling-based DDoS attacks~\cite{bib:leeDetectionDoS3G09,bib:traynorCellularBotnet09,bib:mullineriBot10,bib:mullinerSignaling12}.

Mobile malware uses various infection vectors in order to gain access to the device; the top two main categories of attacks based on the vulnerabilities they use are:
\begin{itemize}
  \item \textit{Exploiting} hardware or software vulnerabilities of devices to completely bypass the user and instal malware. Some of the exploits used to  attack smart devices are near field communication (NFC) technology, third-party kernel drivers, Android firmware vulnerabilities, and mobile web browsers. For example, some malicious websites use mobile browser exploits to install malware on the device without any user interaction other than visiting the site. Android firmware and third-party driver exploits have been used by malware to elevate their privileges and thus gain root access to the device, allowing them to practically do anything they want without the user's knowledge~\cite{bib:liebergeldAndroidSecurity13}.
  
  \item \textit{Social engineering} is by far the most common method used to infect smart mobile devices, where users are ``tricked'' into installing the malware themselves. Social engineering includes all techniques that exploit the human user, such as phishing, application repackaging, etc., in order to infect the device. Social engineering is popular since it does not require any technical investment by the attacker, i.e.~the identification of a new exploit and the development of a delivery system that uses it. Upcoming malware will continue to employ social engineering in new ways; for example, we have already witnessed the first malicious QR codes, which need to be scanned by the user for their activation.
\end{itemize}
Independently of how it infects the device, once the malware is installed, it performs one or more malicious activities, which are classified in Table \ref{tab:malwareBehaviour} according to their behavior. Figure \ref{fig:malwareTypes} shows the distribution of malware types for the top ten Android malware families in 2012, based on data from TrendLabs~\cite{bib:trendmicroMobileThreats12}. The data shows that some malware families exhibit multiple malicious activities (e.g. both premimum service abuse and data stealing), and that premium service abuse has been the favorite malware type, most probably because it is simple to create and allows a direct source of revenue for the cyber-criminal. It is closely followed by adware, which generates indirect profit through advertisement fraud. Malicious downloaders appear to be popular malware delivery methods since they can evade detection by malware detectors and do not alarm the user at installation time by requesting many high-level privileges.

\setlength{\tabcolsep}{8pt}
\renewcommand{\arraystretch}{1.5}

\begin{table}[tbp]
	\caption{Malware behavioral classification.}
	\label{tab:malwareBehaviour}
	\centering
	\begin{tabular}{l p{5cm}}
		\toprule
		Activity & Description\\
		\midrule
		Stealing user information	& Steals user information and credentials. The most commonly targeted data are the contact list, IMEI and IMSI numbers, API authentication keys, banking credentials, user's location, network operator, phone ID and model, phone number, and text messages.\\
		Monitoring					& Tracks user's location, records conversations, takes photos. Spyware is considered to exhibit this type of behavior.\\
		Adware						& Presents unwanted advertisements to the user. Most mobile adware have evolved to incorporate other types of behavior, such as monitoring user activity, especially browsing behavior, and stealing user information. Malicious advertisement networks are increasingly finding their way into legitimate applications and being used as infection vectors.\\
		Premimum service abuse		& Sends premium SMS/MMS, makes premium calls, subscribes the user to premium services without the user's knowledge. Cyber-criminals often make direct financial gain from such premium service abuse.\\
		Click fraud					& Generates ``clicks'' on ads shown on websites and applications, generating indirect financial gain to the cyber-criminal through payment from the advertisement networks.\\
		Search engine optimization	& Manipulates the search results shown on the mobile browser and other applications to improve website rankings in search engine results.\\
		SMS and e-mail spam			& Sends spam SMS and e-mail either to the user's contacts or to a specified list of people. Could be used for phishing attacks.\\
		Malicious downloading		& Downloads malicious content onto the device. Mainly used to evade detection by malware detectors and the user.\\
		Botclient					& Turns the device into a botclient that receives commands from a remote command and control server. Once part of a mobile botnet, the device can be used to launch a variety of attacks, ranging from spam to DDoS attacks on the network.\\
		Rooting						& Roots the device to allow execution of otherwise restricted commands and programs. Malware that has root access potentially has full control of the device.\\
		Ransom						& Locks the device and demands a ransom to be paid in order to unlock the device.\\
		Destruction					& Causes physical damage by deleting important files or personal information.\\
		Denial-of-service			& Launches a denial-of-service attack either on the mobile device itself or on the core network. The mobile device may be attacked by repeatedly switching the device off or depleting the battery.\\
		\bottomrule
	\end{tabular}
\end{table}

\subsection{Malware Detection Techniques}

The traditional approach to malware detection is \textit{signature-based}, where signatures for identified malware are extracted and used to detect new infections of the same malware on other devices. While this can be very effective for controlling an outbreak, it cannot defend against malware unless samples have already been obtained and analyzed. Therefore, new and unknown malware cannot be detected via signature-based approaches. In order to detect previously unknown malware, \textit{behavioral methods} are useful in which the activities of an application are analyzed via either static or dynamic analysis. \textit{Static analysis} provides a quick and efficient way to detect malware without executing them, but they are ultimately limited in their effectiveness unless the source code for the application is available. Unlike static analysis, \textit{dynamic analysis methods} execute the application code in an isolated environment, for example a virtual machine or a sandbox, so its behavior can be directly observed. Dynamic analysis techniques include function call monitoring, function parameter analysis, information flow tracking, and instruction tracing. An overview of mobile malware detection methods is presented in~\cite{bib:chandraMobileMalware12}, and a more comprehensive survey is given by~\cite{bib:egeleMalwareAnalysis12}. In the rest of this section, we will only consider behavior-based detection methods.

Based on where the detection is performed, we can classify malware detection methods into three categories. \textit{Client-side detectors} reside in the mobile device, but are constrained by its limited physical resources, especially battery. \textit{Network-side detection} methods analyze mobile network traffic from many users and offer a complementary means for detecting attacks targeting mobile users. They can be used in conjunction with client-side methods to improve detection rates, and provide a broad view of malicious activities within a carrier's network, enabling detection of anomalous behavior that would not be visible on a per user basis. However, network-based methods are limited in that they can only monitor and analyze mobile traffic that goes through the cellular network, and therefore may not be suitable for certain malware types, such as malware that exclusively uses \mbox{Wi-Fi} for communication. \textit{Cloud-based detection} offers a trade-off between network-level analysis and on-device security by offloading intensive security analysis and computations to the cloud while monitoring internal mobile device events as well as different types of wireless communications from many users. However, cloud-based solutions can only protect those users that install the application and require a large number of subscribers in order to identify large-scale events, while network-based detection does not require the user to do anything as all detection is performed using data available to the network operator.

\section{The NEMESYS Approach}
\label{sec:nemesys}

Despite recent advances in mobile security, large gaps remain in our understanding of the new and future threats and vulnerabilities in the smart mobile ecosystem. Our initial research has identified the following open issues with respect to the general problem of cyber threats against smart mobile devices:

\textit{- Missing infrastructure for collecting attack traces.} Without an infrastructure to collect attack traces against mobile devices, we will not be able to detect, analyze and understand the evolving attack strategies employed by cyber-criminals. These are crucial in order to develop effective mitigation strategies and to provide seamless services in the smart mobile ecosystem.

\textit{- Virtualization.} By leveraging advances in mobile virtualization technology, we can protect mobile users from the consequences of malware by appropriately restricting access to device functionality. Virtualization would also allow an easy way to reset mobile devices to a state before infection.

\textit{- Mobile honeypot development.} Honeypots are successfully used in wired networks in order to study the strategies of attackers and to protect production systems from attacks. However, the development of mobile honeypots is still at an early stage. For example, the mobile honeypot presented in~\cite{bib:songMobileHoneypot12} can connect to the Internet only over Wi-Fi or a wired connection. Although the mobile honeypot presented in~\cite{bib:wahlischMobileHoneypot13} can connect to a UMTS network via a USB dongle, the honeypot uses desktop computers to emulate mobile devices. We aim to address the deficiencies of these early efforts by developing a high-interaction mobile honeypot using real mobile devices.

\textit{- New potential for exploiting security vulnerabilities through mobile botnets.} While botnets are a well-known phenomenon in the wired Internet, we have just witnessed the first mobile botnets in 2012. Mobile botnets pose interesting questions as to their capabilities and uses since smart mobile devices possess many abilities not present on a desktop computer. Such mobile botnets could be used to attack mobile users (e.g. SMS spam) and the core network (e.g. DDoS attack); the ability of a large number of mobile devices to effectively attack the core network has been demonstrated in recent cases of legitimate but poorly-designed mobile applications. We need to explore the new threats posed by the emergence of mobile botnets in more detail.

\textit{- Adaptability of cyber-criminals and rapidly changing cyber-crime tactics.} Cyber-criminals have become adept at modifying their strategies and tactics as methods are developed to counter their activities. Thus, security solutions should rely less on signatures and instead adopt other forms of detection.

\textit{- New types of attacks that cross service, platform and network boundaries.} Identification of anomalous behavior within a large set of heterogeneous data is difficult and time-consuming, particularly across layers. Statistical analysis is further challenged by rare anomalous events in massive amounts of data.

\textit{- Attack attribution and understanding of the cyber-criminals' modus operandi.} The selection of the best mitigation strategy requires understanding of new phenomena and recognizing changes in how the malicious actors operate. This requires that attacks are analyzed in a detailed way, in order to ``attribute'' responsibility to an exact attacker or to protect the true targets.

\begin{figure*}[tbp]
	\centering
	\includegraphics[width=0.8\linewidth]{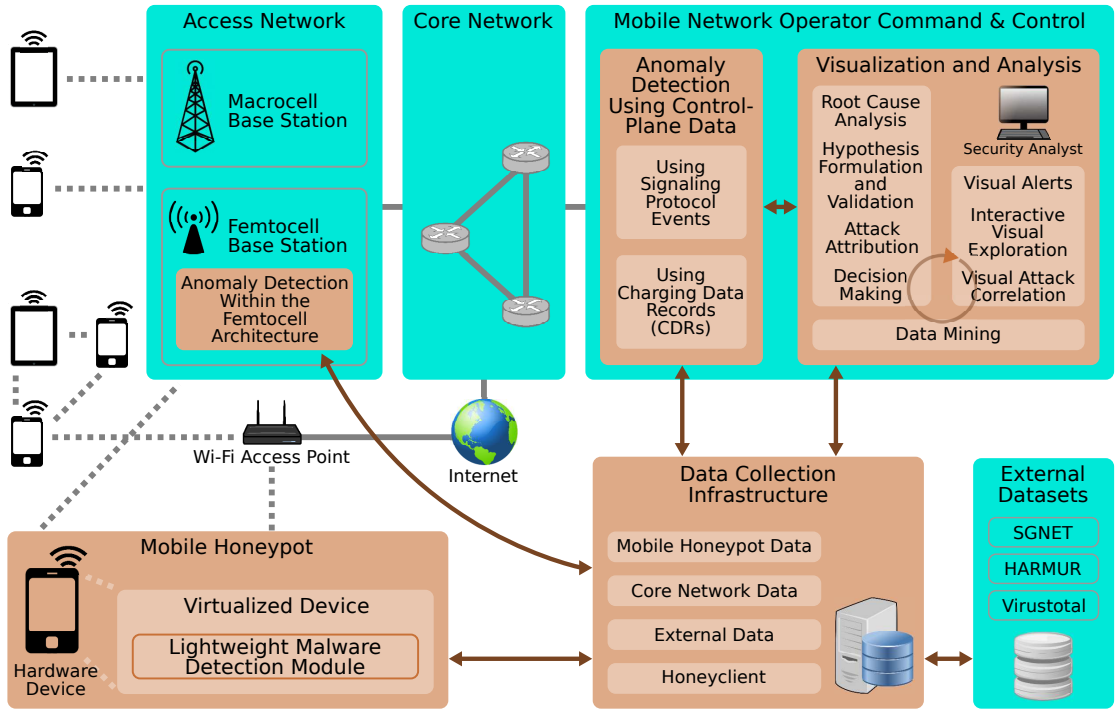}
	\caption{The NEMESYS architecture}
	\label{fig:architecture}
\end{figure*}

In the EU FP7 NEMESYS project, we are developing a data collection, visualization and analysis infrastructure (Fig.~\ref{fig:architecture}) in order to address these open issues. The core of this architecture consists of a data collection infrastructure that incorporates high-interaction virtualized mobile honeypots and honeyclients in order to gather data regarding mobile attacks. The collected data is enriched by the data collection infrastructure through interaction with external sources and made available to anomaly detection, visualization and analysis modules running on the mobile network operator's site. The purpose of the anomaly detection mechanisms is to detect deviations from normal behaviour of mobile users and the core network in real-time. These mechanisms will utilize charging data records (CDRs) for the users and control-plane protocol data, combined with enriched mobile attack traces made available by the data collection infrastructure. In addition to monitoring abnormal behaviour of users connected to the core network through the radio access network, the architecture contains a module that performs anomaly detection within the femtocell complementary access system.

Enriched attack traces and normal behaviour statistics from the data collection infrastructure and the output of the anomaly detection module are fed into the visualization and analysis module. The visualization and analysis module will aid in the detection of existing and emerging threats in the mobile ecosystem through attack attribution, root cause identification, and correlation of observed mobile attacks with known attack patterns. It will present large sets of data related to the observed attacks from heterogeneous sources and facilitate the role of the security analyst in reasoning, hypothesis testing and decision making. We describe the components of the NEMESYS architecture in the following sections in more detail.

\subsection{Virtualized Mobile Client Honeypot}
\label{sec:honeypot}

Honeypots are networked computer system elements that are designed to be attacked and compromised so we can learn about the methods employed by the attackers~\cite{bib:provosHoneypot07}. Traditional honeypots are servers that passively wait to be attacked, whereas client honeypots are security devices that actively search for malware, compromised websites and other forms of attacks. High-interaction client honeypots are fully functional, realistic client systems and they generally do not impose any limitations on the attacker other than those required for containing the attack within the compromised system. Despite their complexity and maintenance difficulty, high-interaction client honeypots are effective at detecting unknown attacks and they are harder to detect by the attacker~\cite{bib:provosHoneypot07}. They also enable in-depth analysis of the attacks during and after the attack has taken place.

As part of NEMESYS, we are developing a high-interaction virtualized client honeypot for the Android mobile platform in order to attract and collect mobile attack traces. We have chosen Android considering its popularity among mobile users and the extremely high ratio of malware targeting Android. Our virtualization technology logically partitions the physical device into two virtual machines (VMs): the \textit{honeypot VM} and the \textit{infrastructure VM}. The honeypot VM will host the largely unmodified mobile device operating system, and it will not have direct access to the device's communication hardware. The infrastructure VM will mediate all access to the communication hardware, and employ sensors to wiretap any communication and detect suspicious behaviour. It will also provide the event monitoring, logging and filesystem snapshot facilities, as well as transmit threat information to the NEMESYS data collection infrastructure. It will host a lightweight malware detection module in order to identify malicious applications running on the honeypot VM. For this purpose, both signature-based and behaviour-based approaches will be considered. In order to improve the efficiency of malware detection, we will identify and prioritize the most important attributes in the system state space to monitor.
Our virtualization technology will ensure that an attack is confined within the compromised device and that it will not put other devices in the network at risk. Furthermore, through this approach, we will be able to stop malware from abusing premium services and from subscribing the user to services without her knowledge. Thus, the user will be spared from any financial distress that may arise as a result of using the mobile honeypot. The virtualization solution also enables taking full snapshots of the honeypot VM filesystem for further forensic analysis of an attack, as well as improving honeypot maintenance since a compromised honeypot could be restored more quickly.

Our initial research has shown that the infection vector of most mobile malware is social engineering, where users are ``tricked'' into installing the malware themselves. This observation has led us to the conclusion that the user should not be ignored in the construction of an effective mobile honeypot. To this end, we introduce the \textit{nomadic honeypot} concept, which utilizes real smartphone hardware with the virtualization solution that will be developed within NEMESYS~\cite{bib:liebergeldHoneypot13}. We plan to deploy nomadic honeypots by handing them out to a chosen group of volunteers, who will use the honeypot as their primary mobile device. It will be up to these human users to get the honeypot infected by visiting malicious sites, installing dubious applications, and so forth. Traces from malware and other types of mobile attacks collected and identified through the nomadic honeypots will be provided to the data collection infrastructure, which is described next.

\subsection{Data Collection Infrastructure}
\label{sec:dataCollection}

The data collection infrastructure will gather and store mobile attack traces that will be provided by the virtualized mobile client honeypot and the honeyclient, and combine them with data from the mobile core network and external sources for enrichment, correlation analysis, and visualization. As an initial step in the design of this infrastructure, we are identifying available external data sources relating to wireline network attacks which will enable correlation of data from multiple heterogeneous sources. Examples of such data sources are the SGNET~\cite{bib:leitaSgnet08}, HARMUR~\cite{bib:leitaHarmur11}, and VirusTotal databases. A source aggregator is being designed and developed to harvest and consolidate data from these sources and the NEMESYS mobile honeypot in a scalable database. Scalable design of the database is important in order to be able to efficiently store and handle large heterogeneous data sets.
Once data from multiple sources have been consolidated, they will be enriched by analyzing the data itself or accessing external sources. For example, TCP/IP stack fingerprinting in order to identify the remote machine's operating system, and clustering of the traces are passive methods of data enrichment. On the other hand, DNS reverse name lookup, route tracing, autonomous system identification, and geo-localization are methods to improve characterization of remote servers but these functions may require access to external sources, possibly in real time. As a final step, the data collection infrastructure will help in the definition of the appropriate inputs representing normal and malicious network activity, which will then be used as the fundamental representation of information in the visualization and analysis module.

The \textit{honeyclient}~\cite{bib:delosieresMalwareDetection13} being developed as part of the data collection infrastructure is similar in concept to the virtualized mobile honeypot, but instead of using real hardware and being driven by real users, the honeyclient uses an Android emulator driven by artificially generated user input to automate interaction with web sites, application markets and applications in order to collect mobile attack traces. The honeyclient consists of a crawler, client, and detector components. The crawler discovers web sites, application markets and applications of interest and generates a list of web pages to visit and applications to download. The client runs the Android emulator (e.g. on a desktop computer) and processes the list generated by the crawler, visiting the web sites using a mobile browser and downloading, installing and executing applications. The behavior of the applications, e.g. function calls, and changes to the system as a result of executing the applications are recorded by the client, which are used by the malware detector to identify malware. The honeyclient provides data relating to identified malware and malicious web sites to the data collection infrastructure.

\subsection{Anomaly Detection Using Control Plane and Billing Data}
\label{sec:anomalyDetection}

The anomaly detection module that operates at the mobile network operator's site is used for the identification and prediction of abnormal behavior observed on smart mobile devices and the mobile network. In addition to user-oriented attacks, mobile networks are vulnerable to a novel DoS attack called the signaling attack~\cite{bib:leeDetectionDoS3G09}. Signaling attacks seek to overload the control plane of the mobile network using low-rate, low-volume attack traffic, based on the structure and characteristics of mobile networks. Unlike conventional DoS attacks that focus on the data plane, the signaling attack creates havoc in the control plane of a mobile network by repeatedly triggering radio channel allocations and revocations. In order to identify such DoS attacks against the mobile network and attacks against the mobile users in real time, we will use signaling data from control-plane protocols and sanitized (anonymized) CDRs from mobile users, respectively. For this purpose, we will use normal user behavior statistics, as well as synthetic ``typical'' user playbacks, to create traces of signaling events and billing data so as to characterize and extract their principal statistics such as frequencies, correlations, times between events, and possible temporal tendencies over short (milliseconds to seconds) and long (hours to days) intervals. We will then employ Bayesian techniques such as maximum likelihood detection, neuronal techniques based on learning, and a combination of these two in order to design and develop robust and accurate change detection algorithms to detect the presence of an attack, and classification algorithms to identify the type of attack when it is detected with high confidence.

Novel femtocell architectures provide a specific opportunity for user-end observation of network usage, while they also have specifics for attacks within the femtocells~\cite{bib:borgaonkarSecurityFemtocell11}. To address attacks specific to femtocells, we will conduct a survey and evaluation of how users may be monitored and attacks detected within a femtocell, and how these are linked to overall mobile network events.

A number of novel ideas are also being investigated~\cite{bib:abdelrahmanAnomalyDetection13} such as modeling the signaling and billing network as a queueing network~\cite{bib:gelenbeMuntzProb76,bib:GelenbeActa} to capture the main events that involve hundreds of thousands of mobile calls and interactions, while only a few may be subject to an intrusion or attack at any given time. Detection of anomalies is studied using learning with neural networks~\cite{bib:gelenbeRNN99,bib:gelenbeNatural12} that provide fast low-order polynomial detection complexity required for massive real-time data, and the need to detect and respond to threats in real-time. Such techniques can also benefit from distributed task decomposition and execution for greater efficiency~\cite{bib:aguilarTask97}.
Our analytical models and anomaly detection algorithms will be augmented and validated with simulation tools. As an initial step, we are developing realistic simulations of UMTS and LTE networks using the OPNET simulator in order to extract data regarding control-plane events that take place during normal mobile communications. Characteristics of these control events will be used to drive the development of our analytical models. We will later conduct large-scale mobile network simulations to validate our mathematical results. Another set of simulations will focus on user-level events, such as voice calls and packet communications, and include charging system components to monitor the use of internal and external network resources. Such simulations will be used to test the performance of our real-time and offline anomaly detection methods.

\subsection{Root Cause Analysis, Correlation and Visualization}
\label{sec:analysisVisualization}

The role of the visualization and analysis module is to process the data obtained from the data collection infrastructure and the anomaly detection module in order to identify and reveal correlations between network events, and to provide a visual analytics framework for the security analyst to perform hypothesis formulation and testing. The data provided to this module represents a large and heterogeneous data set that needs to be presented in a meaningful way to the operator without overwhelming her or restricting available views and actions on the data. In addition to mere representation of data, the visualization and analysis module aims to provide visual analytics tools to the operator. This task is compounded by different uses of visualization by the operator: (i) real-time monitoring of the status of users and the mobile network, and (ii) exploratory data analysis. For real-time monitoring, the security status of a large set of mobile users and more importantly the mobile network need to be presented to the operator. This includes providing early alerts for abnormal behaviour, DoS attacks, malware spreading among the users of the mobile network, etc. The analytics module must also provide visual analytics tools so the analyst can perform attack attribution and correlation analysis with the help of algorithms running in the background.

In order to effectively visualize and explore large sets of heterogeneous, dynamic, complex data, it is necessary to create multiple coordinated views of the data that allow a multi-faceted perception and the discovery of any hidden attributes. The analysis methods also need to be scalable for early network alerts and fast access to the underlying data. We will therefore focus on enabling a real-time analysis framework by means of incremental analysis and visualization methods~\cite{bib:papaVisualNetwork13}, such as multi-level hierarchical screen visualizations that update smoothly rather than showing abrupt changes.

\subsection{Integration and Validation}

In order to evaluate and validate the technologies that are being developed and to demonstrate their impact to interested parties, NEMESYS will construct a virtual testing environment based on guidelines provided by our industrial partners that is as close to a real mobile network as possible within feasibility limitations. The different modules being developed by various partners will be integrated in the virtual testing environment, and validation tests will be conducted based on realistic use-cases. We aim to use the OPNET simulator as part of the virtual testing environment in order to conduct simulations of different types of mobile networks, e.g. UMTS and LTE, and to drive the large-scale networking experiments.

\section{Conclusion}
\label{sec:conclusion}

The evolving and growing nature of the mobile threat in the smart mobile ecosystem is evident from the increasing marketshare of smartphones and tablets, the large amount of data due to smart devices, and the number of detected mobile malware. We must therefore address the mobile threat and understand the new and potential vulnerabilities, threats, and operating methods of cyber-criminals. NEMESYS will provide new insight into the nature of next generation network security in the smart mobile ecosystem. The main innovation of NEMESYS is the research and development of novel security technologies for the identification and prediction of abnormal behavior observed on smart mobile devices, as well as for gathering and analyzing information about the nature of cyber-attacks targeting mobile devices, so that appropriate countermeasures can be taken to combat them. It will involve the development of virtualized honeypots for mobile devices, a data collection infrastructure, and the introduction of novel attack attribution and visual analytics technologies for the mining, presentation and representation of large amounts of heterogeneous data that are related to the smart mobile ecosystem.

\section*{Acknowledgments}

The work presented in this paper was supported by the EU FP7 collaborative research project NEMESYS (Enhanced Network Security for Seamless Service Provisioning in the Smart Mobile Ecosystem), under grant agreement no. 317888 within the FP7-ICT-2011.1.4 Trustworthy ICT domain.


\IEEEtriggeratref{27}

\bibliographystyle{IEEEtran}
\bibliography{references}

\end{document}